\documentclass[prd,reprint,showpacs,showkeys]{revtex4-1}
\usepackage{amssymb}
\usepackage{amsmath}
\usepackage{graphicx}
\usepackage[font={footnotesize,it}]{caption}

\pdfoutput=1

\begin{document}

\title{A new Einstein-nonlinear electrodynamics solution in 2+1-dimensions}
\author{S. Habib Mazharimousavi}
\email{habib.mazhari@emu.edu.tr}
\author{M. Halilsoy}
\email{mustafa.halilsoy@emu.edu.tr}
\author{O. Gurtug}
\email{ozay.gurtug@emu.edu.tr}
\affiliation{Department of Physics, Eastern Mediterranean University, G. Magusa, north
Cyprus, Mersin 10 - Turkey}
\date{\today}

\begin{abstract}
We introduce a class of solutions in $2+1-$dimensional
Einstein-Power-Maxwell  theory for circularly symmetric electric field. The
electromagnetic field is considered with an angular component given by $%
F_{\mu \nu }=E_{0}\delta _{\mu }^{t}\delta _{\nu }^{\theta }$ for $E_{0}=$
constant. First, we show that the metric for zero cosmological constant and
the Power-Maxwell Lagrangian of the form of $\sqrt{\left\vert F_{\mu \nu
}F^{\mu \nu }\right\vert }$, coincides with the solution given in $2+1-$%
dimensional gravity coupled with a massless, self interacting real scalar
field. With the same Lagrangian and a non-zero cosmological constant we
obtain a non-asymptotically flat wormhole solution in $2+1-$dimensions. The
confining motions of massive charged and chargeless particles are
investigated too. Secondly, another interesting solution is given for zero
cosmological constant together with conformal invariant condition. The
formation of timelike naked singularity for this particular case is
investigated within the framework of the quantum mechanics. Quantum fields
obeying the Klein-Gordon and Dirac equations are used to probe the
singularity and test the quantum mechanical status of the singularity.
\end{abstract}

\maketitle

\section{Introduction}

There has always been benefits in studying lower dimensional field
theoretical spacetimes such as $2+1-$dimensions in general relativity. This
is believed to be the projection of higher dimensional cases to the more
tractable situations that may inherit physics of the intricate higher
dimensions. Recent decades proved that the cases of lower dimensions are
still far from being easily understandable and in fact entails its own
characteristics. The absence of gravitational degree of freedom such as Weyl
tensor or pure gravitational waves necessitates endowment of physical
sources to fill the blank and create its own curvatures. Among these the
most popular addition has been a negative cosmological constant which makes
anti-de Sitter spacetimes to the extent that it makes even black holes \cite%
{1}. Addition of electromagnetic \cite{2} and scalar fields \cite{3,4} also
are potential candidates to be considered in the same context. Beside
minimally coupled massless scalar fields which has little significance to
add, non-minimally self-coupled scalar fields has also been considered. In
particular, the real, radial, self-interacting scalar field with a Liouville
potential among others seems promising \cite{4}. The distinctive feature of
the source in such a study is that the radial pressure turns out to be the
only non-zero (i.e., $T_{r}^{r}\neq 0$) component of the energy-momentum
tensor \cite{4}. In effect, such a radial pressure turns out to make a naked
singularity but not a black hole. Being motivated by the self-intracting
scalar field in $2+1-$dimensional gravity we attempt in this paper to do
similar physics with a non-linear electromagnetic field which naturally adds
its own non-linearity. Our choice for the Lagrangian in nonlinear
electrodynamics (NED) is the square root of the Maxwell invariant, i.e. ($%
\sqrt{\left\vert F_{\mu \nu }F^{\mu \nu }\right\vert }$), where as usual the
field tensor is defined by $F_{\mu \nu }=\partial _{\mu }A_{\nu }-\partial
_{\nu }A_{\mu }.$ This Lagrangian belongs to the class of NED with $k-$power
law Maxwell invariant $\left\vert F_{\mu \nu }F^{\mu \nu }\right\vert ^{k}$ 
\cite{5}. Similar Lagrangian in $3+1-$dimensions with magnetic field source
has been considered in \cite{6} and with electric field as well as electric
and magnetic fields together in \cite{7}. A combination of linear Maxwell
with the square root term has been investigated in \cite{8}.

Such a Lagrangian naturally breaks the scale invariance, i.e. $x_{\mu
}\rightarrow \lambda x_{\mu }$, $A_{\mu }\rightarrow \frac{1}{\lambda }%
A_{\mu }$ for $\lambda =$constant, even in $3+1-$dimensions so that
interesting results are expected to ensue. Let us add that such a choice of
NED has the feature that it doesn't attain the familiar linear Maxwell limit
unless $k=1$. One physical consequence beside others, of the square-root
Maxwell Lagrangian is that it gives rise to confinement for geodesic
particles. For a general discussion of confinement in general relativistic
field theory the reader may consult \cite{8}.

Contrary to the previous considerations \cite{4} in this study our electric
field is not radial, instead our field tensor is expressed in the form $%
F_{\mu \nu }=E_{0}\delta _{\mu }^{t}\delta _{\nu }^{\theta }$ for $E_{0}=$%
constant. This amounts to the choice for the electromagnetic vector
potential $A_{\mu }=E_{0}\left( a_{0}\theta ,0,b_{0}t\right) ,$ where our
spacetime coordinates are labelled as $x^{\mu }=\left\{ t,r,\theta \right\} $
and the constants $a_{0}$ and $b_{0}$ satisfy $a_{0}+b_{0}=1.$ The
particular choice $a_{0}=0$, $b_{0}=1$ leaves us with the vector potential $%
A_{\mu }=\delta _{\mu }^{\theta }E_{0}t,$ which yields a uniform field in
the angular direction. The only non vanishing energy momentum tensor
component is $T_{r}^{r}$ which accounts for the radial pressure. Our ansatz
electromagnetic field in the circularly symmetric static metric gives a
solution with zero cosmological constant that is identical with the
spacetime obtained from an entirely different source, namely the
self-intracting real scalar field \cite{4}. This is a conformally flat
anti-de Sitter solution in $2+1-$dimensions without formation of a black
hole. The uniform electric field self-interacting is strong enough to make a
naked singularity at the circular center. Another solution for a non-zero
cosmological constant can be interpreted as a non-asymptotically flat
wormhole solution. In analogy with the $3+1-$dimensions \cite{8} we search
for possible particle confinement in this $2+1-$dimensional model with $%
\Lambda =0.$ Truly the geodesics for both neutral and charged particles are
confined.

In this paper, in addition to provide a new solution in NED theory, we
investigate the resulting spacetime structure for a specific value of $k=3/4,
$ which arises by imposing traceless condition on the Maxwell's energy
momentum, that is known to satisfy conformal invariance condition. In this
particular case the character of the singularity is timelike. For specific
values of the parameters, the timelike character of the naked singularity at 
$r=0$ is also encountered in \cite{12}, in which the radial electric field
is assumed within the context of NED with a power parameter $k=3/4$. This
singularity is investigated within the framework of quantum mechanics in 
\cite{13}. Therein timelike naked singularity is probed with quantum fields
obeying the Klein-Gordon and Dirac equations.

We investigate the timelike naked singularity developed at $r=0,$ for the
new solution, which incorporates a power parameter $k=3/4,$ from the quantum
mechanical point of view. The main motivation to study the singularity is to
clarify whether the uniform electric field in angular direction has an
effect on the resolution of this singularity or not. In order to compare the
present study with the results obtained in \cite{13}, the singularity will
be probed with two different quantum waves having spin structures $0$ and $%
1/2$, namely, the bosonic waves and fermionic waves respectively. The result
of this investigation is that, with respect to the bosonic wave probe the
singularity remains quantum singular, whereas with respect to the fermionic
wave probe the singularity is shown to be healed.

Organization of the paper is as follows. In Sec. II we introduce our
formalism and derive the field equations. New solutions for $k=\frac{1}{2}$
are presented in Sec. III as naked singular / wormhole and discuss its
geodesic confining properties. Sec. IV considers the case with $k=\frac{1}{2}
$ further. The paper ends with Conclusion in Sec. V.

\section{Field equations and the new solution}

We start with the following action for the Einstein's theory of gravity
coupled with a NED Lagrangian 
\begin{equation}
I=\frac{1}{2}\int dx^{3}\sqrt{-g}\left( R-2\Lambda +\alpha \left\vert 
\mathcal{F}\right\vert ^{k}\right) .
\end{equation}%
Here $\mathcal{F}=F_{\mu \nu }F^{\mu \nu }$ is the Maxwell invariant with $%
F_{\mu \nu }=\partial _{\mu }A_{\nu }-\partial _{\nu }A_{\mu }$ , $\alpha $
is a real coupling constant, $k$ is a rational number and $\Lambda $ is the
cosmological constant. Our line element is circularly symmetric given by 
\begin{equation}
ds^{2}=-A\left( r\right) dt^{2}+\frac{1}{B(r)}dr^{2}+r^{2}d\theta ^{2},
\end{equation}%
where $A(r)$ and $B(r)$ are unknown functions of $r$ and $0\leq \theta \leq
2\pi $. Also we choose the field ansatz as 
\begin{equation}
\mathbf{F}=E_{0}dt\wedge d\theta 
\end{equation}%
in which $E_{0}=$constant, is a uniform electric field and its dual can be
found as $^{\star }\mathbf{F}=\frac{E_{0}}{r}\sqrt{\frac{B}{A}}dr.$
Naturally, the integral of $^{\star }\mathbf{F}$ gives the total charge.
This electric field is derived from an electric potential one-form given by%
\begin{equation}
\mathbf{A}=E_{0}\left( a_{0}td\theta -b_{0}\theta dt\right) 
\end{equation}%
in which $a_{0}$ and $b_{0}$ are constants satisfying $a_{0}+b_{0}=1.$ The
nonlinear Maxwell's equation reads%
\begin{equation}
d\left( ^{\star }\mathbf{F}\frac{\left\vert \mathcal{F}\right\vert ^{k}}{%
\mathcal{F}}\right) =0,
\end{equation}%
which upon substitution 
\begin{equation}
\mathcal{F}=2F_{t\theta }F^{t\theta }=\frac{-2E_{0}^{2}}{A\left( r\right)
r^{2}}
\end{equation}%
is trivially satisfied. Next, the Einstein-NED equations are given by 
\begin{equation}
G_{\mu }^{\nu }+\frac{1}{3}\Lambda \delta _{\mu }^{\nu }=T_{\mu }^{\nu }
\end{equation}%
in which 
\begin{equation}
T_{\mu }^{\nu }=\frac{\alpha }{2}\left\vert \mathcal{F}\right\vert
^{k}\left( \delta _{\mu }^{\nu }-\frac{4k\left( F_{\mu \lambda }F^{\nu
\lambda }\right) }{\mathcal{F}}\right) .
\end{equation}%
Having $\mathcal{F}$ known one finds%
\begin{equation}
T_{\ t}^{t}=T_{\ \theta }^{\theta }=\frac{\alpha }{2}\left\vert \mathcal{F}%
\right\vert ^{k}\left( 1-2k\right) ,
\end{equation}%
and 
\begin{equation}
T_{\ r}^{r}=\frac{\alpha }{2}\left\vert \mathcal{F}\right\vert ^{k}
\end{equation}%
as the only non-vanishing energy-momentum component. To proceed further, we
must have the exact form of the Einstein tensor components given by%
\begin{equation}
G_{t}^{t}=\frac{B^{\prime }}{2r}
\end{equation}%
\begin{equation}
G_{r}^{r}=\frac{BA^{\prime }}{2rA}
\end{equation}%
and%
\begin{equation}
G_{\theta }^{\theta }=\frac{2A^{\prime \prime }AB-A^{\prime 2}B+A^{\prime
}B^{\prime }A}{4A^{2}},
\end{equation}%
in which a 'prime' means $\frac{d}{dr}.$ The field equations then read as
follows%
\begin{equation}
\frac{B^{\prime }}{2r}+\frac{1}{3}\Lambda =\frac{\alpha }{2}\left\vert 
\mathcal{F}\right\vert ^{k}\left( 1-2k\right) ,
\end{equation}%
\begin{equation}
\frac{BA^{\prime }}{2rA}+\frac{1}{3}\Lambda =\frac{\alpha }{2}\left\vert 
\mathcal{F}\right\vert ^{k}
\end{equation}%
and%
\begin{equation}
\frac{2A^{\prime \prime }AB-A^{\prime 2}B+A^{\prime }B^{\prime }A}{4A^{2}}+%
\frac{1}{3}\Lambda =\frac{\alpha }{2}\left\vert \mathcal{F}\right\vert
^{k}\left( 1-2k\right) .
\end{equation}

\section{An exact solution for $k=\frac{1}{2}$}

Among the values which $k$ may take the most interesting one is $k=\frac{1}{2%
}.$ In this specific case $T_{\ t}^{t}=T_{\ \theta }^{\theta }=0$ and $T_{\
r}^{r}=\frac{\alpha }{2}\sqrt{\left\vert \mathcal{F}\right\vert }$ is the
only non-zero component of the energy momentum tensor. The field equations
admit the general solutions for $A\left( r\right) $ and $B\left( r\right) $
given by%
\begin{equation}
B\left( r\right) =D-\frac{\Lambda }{3}r^{2}
\end{equation}%
and%
\begin{equation}
A\left( r\right) =\left( D-\frac{\Lambda }{3}r^{2}\right) \left( C+\frac{%
r\alpha \left\vert E_{0}\right\vert }{D\sqrt{2}\sqrt{D-\frac{\Lambda }{3}%
r^{2}}}\right) ^{2}
\end{equation}%
in which $D$ and $C$ are two integration constants. One observes that
setting $E_{0}=0$ gives the correct limit of BTZ black hole up to a constant 
$C^{2}$ which can be absorbed in time. The other interesting limit of the
above solution is found when we set $C=0$ which yields%
\begin{equation}
A\left( r\right) =\left( \frac{r\alpha \left\vert E_{0}\right\vert }{D\sqrt{2%
}}\right) ^{2}.
\end{equation}%
The reduced line element, therefore, becomes%
\begin{equation}
ds^{2}=-\left( \frac{\alpha \left\vert E_{0}\right\vert }{D\sqrt{2}}\right)
^{2}r^{2}dt^{2}+\frac{1}{D-\frac{\Lambda }{3}r^{2}}dr^{2}+r^{2}d\theta ^{2}
\end{equation}%
which for three different cases admits different geometries;

\textbf{i) }$\Lambda >0:$\textbf{\ }When the cosmological constant is
positive the solution becomes non-physical for $r^{2}>\frac{3D}{\Lambda }.$
For $D<0$ the signature of the spacetime is openly violated.

\textbf{ii) }$\Lambda <0:$ In this case the solution is a black string for $%
D>0$ whose Kretschmann scalar is given by 
\begin{equation}
\mathcal{K}=\frac{4\left( r^{4}\Lambda ^{2}-2\Lambda r^{2}D+3D^{2}\right) }{%
3r^{4}},
\end{equation}%
with the singularity at the origin which is also the horizon. However, for $%
D<0$ with negative cosmological constant the solution becomes a wormhole
with the throat located at $r=r_{0}=\sqrt{\frac{3\left\vert D\right\vert }{%
\left\vert \Lambda \right\vert }}$ and 
\begin{equation}
ds^{2}=-\left( \frac{\alpha \left\vert E_{0}\right\vert }{D\sqrt{2}}\right)
^{2}r^{2}dt^{2}+\frac{1}{\left\vert D\right\vert \left( \frac{r^{2}}{%
r_{0}^{2}}-1\right) }dr^{2}+r^{2}d\theta ^{2}.
\end{equation}%
In order to conceive the geometry of the wormhole in this case we introduce
a new coordinate $z=z\left( \rho \right) $ such that%
\begin{equation}
\frac{dr^{2}}{\left\vert D\right\vert \left( \frac{r^{2}}{r_{0}^{2}}%
-1\right) }=dr^{2}+dz^{2}
\end{equation}%
where 
\begin{equation}
z\left( r\right) =\pm \int \left( \sqrt{\frac{1}{\left\vert D\right\vert
\left( \frac{r^{2}}{r_{0}^{2}}-1\right) }-1}\right) dr.
\end{equation}%
It should be supplemented that the ranges of $r$ must satisfy 
\begin{equation}
r_{0}<r.
\end{equation}%
To cast our spacetime into the standard wormhole metric we express it as 
\begin{equation}
ds^{2}=-e^{2f}dt^{2}+\frac{dr^{2}}{\left( 1-\frac{b\left( r\right) }{r}%
\right) }+r^{2}d\theta ^{2}.
\end{equation}%
Here $f\left( r\right) \sim \ln r$ and $b\left( r\right) =r\left(
1+\left\vert D\right\vert -\frac{\left\vert D\right\vert }{r_{0}^{2}}%
r^{2}\right) $ are known as the redshift and shape functions, respectively.
The throat of our wormhole is at $r_{0}$ where $b\left( r_{0}\right) =r_{0},$
and the flare-out condition (i.e., $b^{\prime }\left( r_{0}\right) <1$) is
satisfied by the choice of our parameters. We have also $\frac{b\left(
r\right) }{r}<1$ for $r>r_{0}.$ We must add that distinct from an
asymptotically flat wormhole herein we have a range for $r,$ given in (25).

\textbf{iii) }$\Lambda =0:$ for the case when cosmological constant is zero
one finds (only $D>0$ is physical)%
\begin{equation}
ds^{2}=-\left( \frac{\alpha \left\vert E_{0}\right\vert }{D\sqrt{2}}\right)
^{2}r^{2}dt^{2}+\frac{1}{D}dr^{2}+r^{2}d\theta ^{2}
\end{equation}%
which after a simple rescaling of time and setting $D=1$ one gets%
\begin{equation}
ds^{2}=-r^{2}d\tilde{t}^{2}+dr^{2}+r^{2}d\theta ^{2}.
\end{equation}%
We notice that although our solution is not a standard black hole there
exists still a horizon at $r=0$ which makes our solution a black point \cite%
{9,10}. In \cite{9} such black points appeared in $3+1-$dimensional gravity
coupled to the logarithmic $U(1)$ gauge theory and in \cite{10} coupled to
charged dilatonic fields. This is the conformally flat $2+1-$dimensional
line element, and through the transformation $r=e^{R},$ which is given by 
\begin{equation}
ds^{2}=e^{2R}\left( -d\tilde{t}^{2}+dR^{2}+d\tilde{\theta}^{2}\right)
\end{equation}%
obtained also in the self-intracting scalar field model \cite{4}.

\subsection{Geodesic Motion for $\Lambda =0$}

\subsubsection{Chargeless Particle}

To know more about the solution found above one may study the geodesic
motion of a massive particle (time-like). The Lagrangian of the motion of a
unit mass particle within the spacetime (28) is given by (for simplicity we
remove tildes over the coordinates) 
\begin{equation}
L=-\frac{1}{2}r^{2}\dot{t}^{2}+\frac{1}{2}\dot{r}^{2}+\frac{1}{2}r^{2}\dot{%
\theta}^{2}
\end{equation}%
where a 'dot' denotes derivative $\frac{d}{ds}$ with $s$ an affine
parameter. The conserved quantities are 
\begin{equation}
\frac{\partial L}{\partial \dot{t}}=-r^{2}\dot{t}=-\alpha _{0}
\end{equation}%
\begin{equation}
\frac{\partial L}{\partial \dot{\theta}}=r^{2}\dot{\theta}=\beta _{0}
\end{equation}%
with $\alpha _{0}$ and $\beta _{0}$ as constants of energy and angular
momentum. The metric condition reads 
\begin{equation}
-1=-r^{2}\dot{t}^{2}+\dot{r}^{2}+r^{2}\dot{\theta}^{2}
\end{equation}%
which upon using (31) and (32) yields%
\begin{equation}
\dot{r}^{2}=\left( \frac{\alpha _{0}^{2}-\beta _{0}^{2}}{r^{2}}-1\right) .
\end{equation}%
This equation clearly shows a confinement in the motion for the particle
geodesics in the form%
\begin{equation}
r^{2}\leq \alpha _{0}^{2}-\beta _{0}^{2}.
\end{equation}%
Considering the affine parameter as the proper distance one finds from (41),%
\begin{equation}
r=\sqrt{\alpha _{0}^{2}-\beta _{0}^{2}-\left( s-s_{0}\right) ^{2}}
\end{equation}%
leading to a manifest confinement.

\subsubsection{Charged Particle Geodesics}

For a massive charged particle with unit mass and charge $q_{0}$ the
Lagrangian is given by%
\begin{equation}
L=-\frac{1}{2}r^{2}\dot{t}^{2}+\frac{1}{2}\dot{r}^{2}+\frac{1}{2}r^{2}\dot{%
\theta}^{2}+q_{0}A_{\mu }\dot{x}^{\mu }
\end{equation}%
in which $A_{\mu }\dot{x}^{\mu }=A_{\theta }\dot{x}^{\theta }=E_{0}t\dot{%
\theta}$ i.e. the choice $a_{0}=1,$ $b_{0}=0$ in Eq. (4). The metric
condition is as (33) and therefore the Lagrange equations yield%
\begin{equation}
\frac{d}{ds}\left( r^{2}\dot{\theta}+q_{0}E_{0}t\right) =0,
\end{equation}%
\begin{equation}
\frac{d}{ds}\left( r^{2}\dot{t}\right) =-q_{0}E_{0}\dot{\theta}
\end{equation}%
and 
\begin{equation}
\ddot{r}=-r\dot{t}^{2}+r\dot{\theta}^{2}.
\end{equation}%
The first equation implies%
\begin{equation}
r^{2}\dot{\theta}+q_{0}E_{0}t=\gamma _{0}=const.
\end{equation}%
while the second equation with a change of variable as $r^{2}\frac{d}{ds}=%
\frac{d}{dz}$ and imposing (41) yields%
\begin{equation}
\frac{d^{2}t}{dz^{2}}=-q_{0}E_{0}\left( \gamma _{0}-q_{0}E_{0}t\right) .
\end{equation}%
This equation admits an exact solution for $t\left( z\right) $%
\begin{equation}
t\left( z\right) =\frac{\gamma _{0}}{\omega }+C_{1}e^{\omega
z}+C_{2}e^{-\omega z}
\end{equation}%
in which $C_{1}$ and $C_{2}$ are two integration constants and $\omega
=q_{0}E_{0}.$ Next, the radial equation upon imposing the metric condition
(33) becomes decoupled as%
\begin{equation}
r\ddot{r}+\dot{r}^{2}+1=0.
\end{equation}%
The general solution for this equation is given by%
\begin{equation}
r=\pm \sqrt{\tilde{C}_{1}+2\tilde{C}_{2}s-s^{2}}
\end{equation}%
where we consider the positive root. Once more going one finds%
\begin{equation}
r^{2}\frac{dt}{ds}=\frac{dt}{dz}
\end{equation}%
which in turn becomes%
\begin{equation}
\left( \tilde{C}_{1}+2\tilde{C}_{2}s-s^{2}\right) \frac{dt}{ds}=\omega
\left( C_{1}e^{\omega z}-C_{2}e^{-\omega z}\right) .
\end{equation}%
To proceed further we set $\tilde{C}_{2}=0,\tilde{C}_{1}=b_{0}^{2},$ $C_{2}=0
$ and $C_{1}=1$ so that (one must note that with this choice of integration
constants $b_{0}^{2}-s^{2}\geq 0$) 
\begin{equation}
\left( b_{0}^{2}-s^{2}\right) \frac{dt}{ds}=\omega \left( t-\frac{\gamma _{0}%
}{\omega }\right) .
\end{equation}%
This leads us to%
\begin{equation}
\left\vert t-\frac{\beta _{0}}{\omega }\right\vert =\left( \frac{b_{0}+s}{%
b_{0}-s}\right) ^{\frac{\omega }{2b_{0}}}+\zeta ,
\end{equation}%
in which $\zeta $ is an integration constant to be set to zero for
simplicity. Having the latter relation one finds 
\begin{equation}
r\left( t\right) =\frac{2b_{0}\left\vert t-\frac{\gamma _{0}}{\omega }%
\right\vert ^{\frac{b_{0}}{\omega }}}{\left\vert t-\frac{\gamma _{0}}{\omega 
}\right\vert ^{\frac{2b_{0}}{\omega }}+1}
\end{equation}%
which clearly shows a confining motion for the charged particles. This
conclusion could also be obtained from Eq. (45) which implies $\tilde{C}%
_{1}+2\tilde{C}_{2}s-s^{2}\geq 0$ and consequently $r\leq \sqrt{\tilde{C}%
_{1}+\tilde{C}_{2}^{2}}.$ The angular variable $\theta \left( t\right) $ can
also be reduced to an integral expression.

\section{A brief Account for a conformally invariant Maxwell source with $%
\Lambda =0$}

In the first paper of Hassa\"{\i}ne and Mart\'{\i}nez \cite{5} it was shown
that the action given in (1) is conformally invariant if $k=\frac{3}{4}.$
Here in this section we set $k=\frac{3}{4}$ and $\Lambda =0$ such that the
following solution is obtained for the field equations (14)-(16):%
\begin{equation}
A\left( r\right) =\left( 1+\frac{\alpha \left( E_{0}^{2}\right) ^{3/4}}{%
2^{1/4}}\sqrt{r}\right) ^{4}
\end{equation}%
and%
\begin{equation}
B(r)=\frac{1}{\left( 1+\frac{\alpha \left( E_{0}^{2}\right) ^{3/4}}{2^{1/4}}%
\sqrt{r}\right) ^{2}}.
\end{equation}%
To find these solutions we have to set the integration constant in such a
way that the flat limit easily comes after one imposes $E_{0}=0.$ The
Kretschmann scalar of the spacetime is given by%
\begin{equation}
K=\frac{3\left\vert E_{0}\right\vert ^{3}\alpha ^{2}\left( r\alpha
^{2}\left\vert E_{0}\right\vert ^{3}+2\sqrt[4]{2}\left( E_{0}^{2}\right)
^{3/4}\sqrt{r}\alpha +\sqrt{2}\right) }{r^{3}\left( 1+\frac{\alpha \left(
E_{0}^{2}\right) ^{3/4}}{2^{1/4}}\sqrt{r}\right) ^{8}}
\end{equation}%
As one can see from the action (1) $\alpha <0$ implies a ghost field which
is not physical. Therefore we assume $\alpha >0$ and consequently the only
singularity of the spacetime is located at the origin $r=0.$ For $\alpha >0$
the solution represents a non-black hole and non-asymptotically flat
spacetime with a naked singularity at $r=0$. In order to find the nature or
the character of the singularity at $r=0$, we perform a conformal
compactification. The conformal radial / tortoise coordinate 
\begin{equation}
r_{\ast }=\int \frac{dr}{1+a\sqrt{r}}=\frac{2}{a^{2}}\left\{ a\sqrt{r}-\ln
\left\vert 1+a\sqrt{r}\right\vert \right\} ,
\end{equation}%
with $a=$ $\frac{\alpha \left( E_{0}^{2}\right) ^{3/4}}{2^{1/4}}$ helps us
to introduce the retarded and advanced coordinates i.e. $u=t-r_{\ast }$ and $%
v=t+r_{\ast }$. The Kruskal coordinates are defined following $u$ and $v$
coordinates as 
\begin{equation}
u^{^{\prime }}=-e^{a^{2}u},\text{ \ \ \ \ \ \ \ }v^{^{\prime }}=e^{-a^{2}v}%
\text{\ ,}
\end{equation}%
which consequently the line element in $u^{\prime }$ and $v^{\prime }$
coordinates becomes%
\begin{equation}
ds^{2}=-\frac{e^{4a\sqrt{r}}}{a^{4}}du^{^{\prime }}dv^{^{\prime
}}+r^{2}d\theta ^{2},
\end{equation}%
with the constraint 
\begin{equation}
u^{^{\prime }}v^{^{\prime }}=-e^{-4a\sqrt{r}}\left( 1+a\sqrt{r}\right) ^{4}.
\end{equation}%
The final change of the coordinate 
\begin{eqnarray}
u^{^{\prime \prime }} &=&\arctan u^{^{\prime }},\text{ \ \ \ \ \ \ }%
0<u^{^{\prime \prime }}<\pi /2,\text{\ } \\
\text{\ \ \ \ \ \ }v^{^{\prime \prime }} &=&\arctan v^{^{\prime }},\text{ \
\ \ \ \ \ }0<v^{^{\prime \prime }}<\pi /2  \notag
\end{eqnarray}%
brings infinity into a finite coordinate. In Fig. 1 we plot Carter-Penrose
diagrams of the solutions (51) and (52). One observes in this diagram that
the singularity located at $r=0$ is timelike. The corresponding energy
momentum tensor%
\begin{equation}
T_{\mu }^{\nu }=-\frac{1}{2}\alpha \left( \frac{2E_{0}^{2}}{A\left( r\right)
r^{2}}\right) ^{3/4}\left( 1,-1,1\right) 
\end{equation}%
implies that 
\begin{eqnarray}
\rho  &=&-T_{t}^{t}=\frac{1}{2}\alpha \left( \frac{2E_{0}^{2}}{A\left(
r\right) r^{2}}\right) ^{3/4},\ \  \\
p &=&T_{r}^{r}=\rho =\frac{1}{2}\alpha \left( \frac{2E_{0}^{2}}{A\left(
r\right) r^{2}}\right) ^{3/4}, \\
\text{\ }q &=&T_{\theta }^{\theta }=-\rho =-\frac{1}{2}\alpha \left( \frac{%
2E_{0}^{2}}{A\left( r\right) r^{2}}\right) ^{3/4}
\end{eqnarray}%
and therefore all energy conditions including the weak, strong and dominant
are satisfied. 
\begin{figure}[tbp]
\includegraphics[width=30mm,scale=0.7]{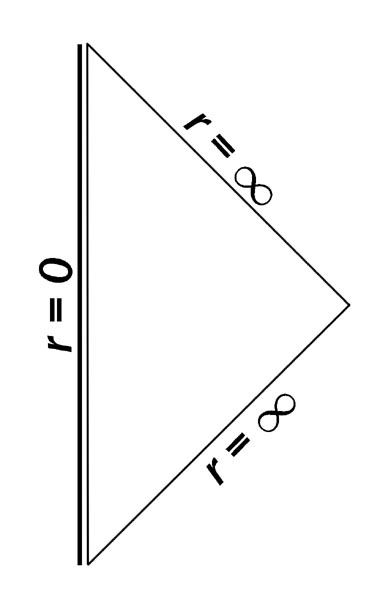}
\caption{Carter-Penrose diagram for $k=\frac{3}{4}$ and $\Lambda =0$ i.e.
Eqs. (51) and (52). We see that $r=0$ is a naked timelike singularity. }
\end{figure}

\section{Singularity Analysis}

\subsection{Quantum Singularities}

One of the important predictions of the Einstein's theory of relativity is
the formation of the spacetime singularities in which the evolution of
timelike or null geodesics is not defined after a proper time. Deterministic
nature of general relativity requires that the spacetime singularities must
be hidden by horizon(s), as conjectured by Penrose's weak cosmic censorship
hypothesis (CCH). However, there are some cases that the spacetime
singularity is not covered by horizon(s), and it is called naked
singularity. Hence, naked singularities violates the CCH and their
resolution becomes extremly important. The most powerful candidate theory in
resolving the singularities is the quantum theory of gravity. However, there
is no consistent quantum theory of gravity yet. String theory \cite{14,15}
and loop quantum gravity \cite{16} are the two study field along the
resolution of singularities. Another method that we shall employ in this
paper is the criterion proposed by Horowitz and Marolf (HM) \cite{17} that
incorporates "self - adjointness" of the spatial part of the wave operator.
In this criterion, the classical notion of \textit{geodesics incompletness}
with respect to point-particle probe will be replaced by the notion of 
\textit{quantum singularity} with respect to wave probes. This criterion can
be applied only to static spacetimes having timelike singularities. To
understand better, let us consider, the Klein-Gordon equation for a free
particle that satisfies $i\frac{d\psi }{dt}=\sqrt{\mathcal{A}_{E}}\psi ,$
whose solution is $\psi \left( t\right) =\exp \left[ -it\sqrt{\mathcal{A}_{E}%
}\right] \psi \left( 0\right) $ in which $\mathcal{A}_{E}$ denotes the
extension of the spatial part of the wave operator.

If $\mathcal{A}$ is not essentially self-adjoint, in other words if $%
\mathcal{A}$ has an extension, the future time evolution of the wave
function $\psi \left( t\right) $ is ambiguous. Then the HM criterion defines
the spacetime as quantum mechanically singular. However, if there is only a
single self-adjoint extension, the operator $\mathcal{A}$ is said to be\
essentially self-adjoint and the quantum evolution described by $\psi \left(
t\right) $ is uniquely determined by the initial conditions. According to
the HM criterion, this spacetime is said to be quantum mechanically
non-singular. The essential self-adjointness of the operator $\mathcal{A}$,
can be verified by considering solutions of the equation%
\begin{equation}
\mathcal{A}^{\ast }\psi \pm i\psi =0,
\end{equation}%
and showing that the solutions of Eq. (63), do not belong to Hilbert space $%
\mathcal{H}$ (we refer references; \cite{18} for detailed mathematical
analysis and \cite{19} for applications of HM approach in different
spacetimes). This will be achieved by defining the function space on each $%
t= $constant hypersurface $\Sigma $ as $\mathcal{H}=\{\left. R\right\vert
\left\Vert R\right\Vert <\infty \}$ with the following norm given for the
metric (2) as,%
\begin{equation}
\left\Vert R\right\Vert ^{2}=\int_{0}^{\text{constant}}\frac{r}{\sqrt{%
A(r)B(r)}}\left\vert R\right\vert ^{2}dr.
\end{equation}

\subsubsection{Klein-Gordon Fields}

Klein-Gordon equation for the metric (2) can be written by splitting
temporal and spatial part and given by%
\begin{equation}
\frac{\partial ^{2}\psi }{\partial t^{2}}=-\mathcal{A}\psi
\end{equation}%
in which $\mathcal{A}$ denotes the spatial operator of the massless scalar
wave given by%
\begin{multline}
\mathcal{A}=-\left( 1+a\sqrt{r}\right) ^{2}\frac{\partial ^{2}}{\partial
r^{2}}-\frac{\left( 1+a\sqrt{r}\right) \left( 1+\frac{3a\sqrt{r}}{2}\right) 
}{r}\frac{\partial }{\partial r} \\
-\frac{\left( 1+a\sqrt{r}\right) ^{4}}{r^{2}}\frac{\partial ^{2}}{\partial
\theta ^{2}}.
\end{multline}%
\ Applying separation of variables $\psi =R(r)Y(\theta ),$ the radial part
of the Eq. (63) becomes%
\begin{multline}
\frac{\partial ^{2}R(r)}{\partial r^{2}}+\frac{\left( 1+\frac{3a\sqrt{r}}{2}%
\right) }{r\left( 1+a\sqrt{r}\right) }\frac{\partial R(r)}{\partial r}+ \\
\left( \frac{c\left( 1+a\sqrt{r}\right) ^{2}}{r^{2}}\pm \frac{i}{\left( 1+a%
\sqrt{r}\right) }\right) R(r)=0,
\end{multline}%
where $c$ stands for the separation constant. The spatial operator $\mathcal{%
A}$ is esentially self adjoint if neither of two solutions of Eq. (67) is
square integrable over all space $L^{2}(0,\infty )$. Because of the
complexity in finding exact analytic solution to Eq. (67), we study the
behavior of $R\left( r\right) $ near $r\rightarrow 0$ and $r\rightarrow
\infty .$ The behavior of the Eq. (67), near $r=0$ is given by%
\begin{equation}
\frac{\partial ^{2}R(r)}{\partial r^{2}}+\frac{1}{r}\frac{\partial R(r)}{%
\partial r}+\frac{c}{r^{2}}R(r)=0,
\end{equation}%
whose solution is%
\begin{equation}
R\left( r\right) =C_{1}\sin \left( \sqrt{c}\ln \left( r\right) \right)
+C_{2}\cos \left( \sqrt{c}\ln \left( r\right) \right)
\end{equation}%
The square integrability is checked by calculating the norm given in Eq.
(64). Calculation has shown that $R\left( r\right) $ is square integrable
near $r=0$ and hence, it belongs to the Hilbert space and the operator $%
\mathcal{A}$ is not essentially self-adjoint. As a result, the timelike
naked singularity remains quantum mechanically singular with respect to the
spinless wave probe. This result is in conform with the analysis in \cite{13}%
. This result seem to show that irrespective of the direction of electric
field, the singularity remains quantum singular with respect to waves
obeying the Klein-Gordon equation.

\subsubsection{Dirac Fields}

The Dirac equation in $2+1$ dimensional curved spacetime for a free particle
with mass $m$ is given by,

\begin{equation}
i\sigma ^{\mu }\left( x\right) \left[ \partial _{\mu }-\Gamma _{\mu }\left(
x\right) \right] \Psi \left( x\right) =m\Psi \left( x\right) ,
\end{equation}%
where $\Gamma _{\mu }\left( x\right) $\ is the spinorial affine connection
given by

\begin{equation}
\Gamma _{\mu }\left( x\right) =\frac{1}{4}g_{\lambda \alpha }\left[ e_{\nu
,\mu }^{\left( i\right) }(x)e_{\left( i\right) }^{\alpha }(x)-\Gamma _{\nu
\mu }^{\alpha }\left( x\right) \right] s^{\lambda \nu }(x),
\end{equation}%
with

\begin{equation}
s^{\lambda \nu }(x)=\frac{1}{2}\left[ \sigma ^{\lambda }\left( x\right)
,\sigma ^{\nu }\left( x\right) \right] .
\end{equation}

Since the fermions have only one spin polarization in $2+1$ dimension \cite%
{20}, the Dirac matrices $\gamma ^{\left( j\right) }$ can be expressed in
terms of Pauli spin matrices $\sigma ^{\left( i\right) }$ \cite{21} so that

\begin{equation}
\gamma ^{\left( j\right) }=\left( \sigma ^{\left( 3\right) },i\sigma
^{\left( 1\right) },i\sigma ^{\left( 2\right) }\right) ,
\end{equation}%
where the Latin indices represent internal (local) frame. In this way,

\begin{equation}
\left\{ \gamma ^{\left( i\right) },\gamma ^{\left( j\right) }\right\} =2\eta
^{\left( ij\right) }I_{2\times 2},
\end{equation}%
where $\eta ^{\left( ij\right) }$\ is the Minkowski metric in $2+1$
dimension and $I_{2\times 2}$\ is the identity matrix. The coordinate
dependent metric tensor $g_{\mu \nu }\left( x\right) $\ and matrices $\sigma
^{\mu }\left( x\right) $\ are related to the triads $e_{\mu }^{\left(
i\right) }\left( x\right) $\ by

\begin{align}
g_{\mu \nu }\left( x\right) & =e_{\mu }^{\left( i\right) }\left( x\right)
e_{\nu }^{\left( j\right) }\left( x\right) \eta _{\left( ij\right) }, \\
\sigma ^{\mu }\left( x\right) & =e_{\left( i\right) }^{\mu }\gamma ^{\left(
i\right) },  \notag
\end{align}%
where $\mu $\ and $\nu $\ stand for the external (global) indices. The
suitable triads for the metric () are given by,

\begin{equation}
e_{\mu }^{\left( i\right) }\left( t,y,\theta \right) =diag\left( \left( 1+a%
\sqrt{r}\right) ^{2},\left( 1+a\sqrt{r}\right) ,r\right) ,
\end{equation}%
The coordinate dependent gamma matrices and the spinorial affine connection
are given by

\begin{align}
\sigma ^{\mu }\left( x\right) & =\left( \sigma ^{\left( 3\right) }\left( 1+a%
\sqrt{r}\right) ^{-2},i\left( 1+a\sqrt{r}\right) ^{-1}\sigma ^{\left(
1\right) },\frac{i\sigma ^{\left( 2\right) }}{r}\right) , \\
\Gamma _{\mu }\left( x\right) & =\left( \frac{a\sigma ^{\left( 2\right) }}{2%
\sqrt{r}},0,0\right) .  \notag
\end{align}%
Now, for the spinor

\begin{equation}
\Psi =\left( 
\begin{array}{c}
\psi _{1} \\ 
\psi _{2}%
\end{array}%
\right) ,
\end{equation}%
the Dirac equation can be written as%
\begin{multline}
\frac{i}{\left( 1+a\sqrt{r}\right) ^{2}}\frac{\partial \psi _{1}}{\partial t}%
-\frac{ai}{2\sqrt{r}\left( 1+a\sqrt{r}\right) ^{2}}\psi _{2}- \\
\frac{1}{\left( 1+a\sqrt{r}\right) }\frac{\partial \psi _{2}}{\partial r}+%
\frac{1}{r}\frac{\partial \psi _{2}}{\partial \theta }-m\psi _{1}=0
\end{multline}

\begin{multline}
\frac{-i}{\left( 1+a\sqrt{r}\right) ^{2}}\frac{\partial \psi _{2}}{\partial t%
}-\frac{ai}{2\sqrt{r}\left( 1+a\sqrt{r}\right) ^{2}}\psi _{1}- \\
\frac{1}{\left( 1+a\sqrt{r}\right) }\frac{\partial \psi _{1}}{\partial r}-%
\frac{1}{r}\frac{\partial \psi _{1}}{\partial \theta }-m\psi _{2}=0
\end{multline}%
The following ansatz will be employed for the positive frequency solutions:

\begin{equation}
\Psi _{n,E}\left( t,x\right) =\left( 
\begin{array}{c}
R_{1n}(r) \\ 
R_{2n}(r)e^{i\theta }%
\end{array}%
\right) e^{in\theta }e^{-iEt}.
\end{equation}%
The radial part of the Dirac equation becomes,%
\begin{multline}
\frac{\partial R_{2n}(r)}{\partial r}-e^{-i\theta }\left( \frac{E}{1+a\sqrt{r%
}}-m\left( 1+a\sqrt{r}\right) \right) R_{1n}(r)- \\
i\left[ \frac{\left( n+1\right) \left( 1+a\sqrt{r}\right) }{r}-\frac{c}{2%
\sqrt{r}\left( 1+a\sqrt{r}\right) }\right] R_{2n}(r)=0,
\end{multline}%
\begin{multline}
\frac{\partial R_{1n}(r)}{\partial r}-e^{i\theta }\left( \frac{1}{1+a\sqrt{r}%
}-m\left( 1+a\sqrt{r}\right) \right) R_{2n}(r)+ \\
i\left[ \frac{n\left( 1+a\sqrt{r}\right) }{r}+\frac{c}{2\sqrt{r}\left( 1+a%
\sqrt{r}\right) }\right] R_{1n}(r)=0.
\end{multline}%
The behavior of the Dirac equations near $r=0$ reduces to%
\begin{multline}
\frac{\partial ^{2}R_{2n}(r)}{\partial r^{2}}+\frac{i}{r}\frac{\partial
R_{2n}(r)}{\partial r}+ \\
\left\{ \frac{n+1}{r^{2}}\left( n+i\right) -\beta \right\} R_{2n}(r)=0,
\end{multline}%
\begin{multline}
\frac{\partial ^{2}R_{1n}(r)}{\partial r^{2}}+\frac{i}{r}\frac{\partial
R_{1n}(r)}{\partial r}+ \\
\left\{ \frac{n}{r^{2}}\left( n+1-i\right) -\beta \right\} R_{1n}(r)=0,
\end{multline}%
in which $\beta =m^{2}-m(E+1)+E.$ These two equations (84) and (85), must be
investigated for essential self-adjointness by using the Eq. (63). Hence, we
have%
\begin{multline}
\frac{\partial ^{2}R_{2n}(r)}{\partial r^{2}}+\frac{i}{r}\frac{\partial
R_{2n}(r)}{\partial r}+ \\
\left\{ \frac{n+1}{r^{2}}\left( n+i\right) -\beta \pm i\right\} R_{2n}(r)=0,
\end{multline}%
\begin{multline}
\frac{\partial ^{2}R_{1n}(r)}{\partial r^{2}}+\frac{i}{r}\frac{\partial
R_{1n}(r)}{\partial r}+ \\
\left\{ \frac{n}{r^{2}}\left( n+1-i\right) -\beta \pm i\right\} R_{1n}(r)=0.
\end{multline}%
Since we are looking for a solution near $r=0$, the constant terms inside
the curly bracket can be ignored and the solutions are given by%
\begin{equation}
R_{2n}(r)=r^{\frac{1-I}{2}}\left[ C_{1n}r^{\frac{\sqrt{-2I-4\chi _{1}}}{2}%
}+C_{2n}r^{\frac{-\sqrt{-2I-4\chi _{1}}}{2}}\right] ,
\end{equation}%
\begin{equation}
R_{1n}(r)=r^{\frac{1-I}{2}}\left[ C_{3n}r^{\frac{\sqrt{-2I-4\chi _{2}}}{2}%
}+C_{4n}r^{\frac{-\sqrt{-2I-4\chi _{2}}}{2}}\right] ,
\end{equation}%
in which $C_{in}$ are arbitrary integration constants 
\begin{equation}
\chi _{1}=\left( n+1\right) \left( n+i\right) ,
\end{equation}%
and%
\begin{equation}
\chi _{2}=n\left( n+1-i\right) .
\end{equation}%
The square integrability of these solutions are checked by using the norm
defined in Eq. (64). Based on the numerical calculation, both $R_{1n}(r)$
and $R_{2n}(r)$ are square integrable near $r=0.$ As a result, the arbitrary
wave packet can be written as%
\begin{equation}
\Psi \left( t,x\right) =\sum_{n=-\infty }^{+\infty }\left( 
\begin{array}{c}
R_{1n}(r) \\ 
R_{2n}(r)e^{i\theta }%
\end{array}%
\right) e^{in\theta }e^{-iEt},
\end{equation}%
and the initial condition $\Psi \left( 0,x\right) $ is enough to determine
the time evolution of the wave. Hence, the initial value problem is
well-posed and the spacetime remains nonsingular when probed with spinorial
waves obeying the Dirac equation.

\section{ Conclusion}

We considered a specific form of NED Lagrangian in the form of a power law
Maxwell invariant $\left\vert F_{\mu \nu }F^{\mu \nu }\right\vert ^{k}$ with 
$k=\frac{1}{2}.$ It is known that a pure radial electric field with $k=\frac{%
1}{2}$ does not satisfy the energy conditions \cite{11} so our field ansatz
has been chosen differently to be a uniform angular electric field. One
direct feature of this form of field ansatz yields the only non-zero
component of the energy-momentum to be $T_{r}^{r}.$ This indeed means that
the energy density $\rho =-T_{t}^{t}$ and the angular pressure $p_{\theta
}=T_{\theta }^{\theta }$ are both zero while the radial pressure is $%
p_{r}=T_{r}^{r}=\frac{\xi }{r}$ with $\xi =\sqrt{\frac{\sqrt{2}\alpha E_{0}}{%
8}}.$ These make the weak energy conditions (WECs) to be satisfied, i.e., $%
\rho \geq 0,$ $\rho +p_{r}\geq 0$ and $\rho +p_{\theta }\geq 0$. Even
further, the strong energy conditions (SECs) are also satisfied which are
WECs together with $\rho +p_{r}+p_{\theta }\geq 0.$ Having radial pressure
non-zero and divergence at $r=0$ are features that can be seen from the
nature of the resulting spacetime. For $k=\frac{1}{2}$ we give two different
classes of solutions, for $\Lambda =0$ and $\Lambda \neq 0.$ The one for $%
\Lambda =0$ is not a black hole solution but is singular at $r=0$ since the
spacetime invariants are $R\sim \frac{1}{r^{2}},$ $R_{\mu \nu }R^{\mu \nu
}\sim \frac{1}{r^{4}}$ and Kretschmann scalar$\sim \frac{1}{r^{4}}$. This
singularity is in the same order of divergence as the charged BTZ black hole
with a radial singular electric field. As it has been found in this work,
the singularity at the origin ($r=0$ is also a zero for $g_{tt}$ which makes
our solution a black point) confines the radial motion of a massive particle
(both charged and uncharged). This means that the particle can not go beyond
a maximum radius. In the work of Schmidt and Singleton \cite{4} where a
different matter source, namely a self-interacting scalar field is employed,
the same solution has been found. This is perhaps an indication that the
geometry found in the context of a real scalar field, sharing common metric
with different sources in $2+1-$dimensions may apply to higher dimensional
spacetimes. The case for $\Lambda <0$ ($k=\frac{1}{2}$) corresponds to a
non-asymptotically flat wormhole built entirely from the cosmological
constant. Conformally invariant case ($k=\frac{3}{4}$) for the Maxwell field
has also been considered briefly. Our timelike null singularity turns out to
be in accordance with quantum regular / singular against the Dirac /
Klein-Gordon probes, respectively.

\end{document}